\documentclass[%
 reprint,
superscriptaddress,
 amsmath,amssymb,
 aps,
 prl,
 longbibliography
]{revtex4-1}

\usepackage{graphicx}
\usepackage{siunitx}
\usepackage{xcolor}
\usepackage{subcaption}
\usepackage[utf8x]{inputenc}
\usepackage{lineno}
\usepackage{supertabular}
\usepackage{hyperref}
\usepackage{soul}
\usepackage{appendix}
\usepackage{ulem}
\usepackage{url}

\definecolor{gesfred}{rgb}{1,0,0}

\makeatletter
\renewcommand\@makecaption[2]{%
  \par
  \vskip\abovecaptionskip
  \begingroup
   \small\rmfamily
    \begingroup
     \samepage
     \flushing
     \let\footnote\@footnotemark@gobble
     \@make@capt@title{#1}{#2}\par
    \endgroup
  \endgroup
  \vskip\belowcaptionskip
}
\makeatother

\begin{document}
\title{Search for light dark matter from atmosphere in PandaX-4T}


\def\shKeyLab{School of Physics and Astronomy, Shanghai Jiao Tong University, Key Laboratory for Particle Astrophysics and Cosmology (MoE), Shanghai Key Laboratory for Particle Physics and Cosmology, Shanghai 200240, China}
\def\BUAA{School of Physics, Beihang University, Beijing 102206, China}
\def\BUAALab{Beijing Key Laboratory of Advanced Nuclear Materials and Physics, Beihang University, Beijing, 102206, China}
\def\zzu{School of Physics and Microelectronics, Zhengzhou University, Zhengzhou, Henan 450001, China}
\def\USTClab{State Key Laboratory of Particle Detection and Electronics, University of Science and Technology of China, Hefei 230026, China}
\def\USTCdep{Department of Modern Physics, University of Science and Technology of China, Hefei 230026, China}
\def\BUAALab{International Research Center for Nuclei and Particles in the Cosmos \& Beijing Key Laboratory of Advanced Nuclear Materials and Physics, Beihang University, Beijing 100191, China}
\def\pku{School of Physics, Peking University, Beijing 100871, China}
\def\YaLongSD{Yalong River Hydropower Development Company, Ltd., 288 Shuanglin Road, Chengdu 610051, China}
\def\IAP{Shanghai Institute of Applied Physics, Chinese Academy of Sciences, 201800 Shanghai, China}
\def\CHEPpku{Center for High Energy Physics, Peking University, Beijing 100871, China}
\def\SDUdep{Research Center for Particle Science and Technology, Institute of Frontier and Interdisciplinary Science, Shandong University, Qingdao 266237, Shandong, China}
\def\SDUlab{Key Laboratory of Particle Physics and Particle Irradiation of Ministry of Education, Shandong University, Qingdao 266237, Shandong, China}
\def\UMD{Department of Physics, University of Maryland, College Park, Maryland 20742, USA}
\def\TDLee{Tsung-Dao Lee Institute, Shanghai Jiao Tong University, Shanghai, 200240, China}
\def\MESJTU{School of Mechanical Engineering, Shanghai Jiao Tong University, Shanghai 200240, China}
\def\SYU{School of Physics, Sun Yat-Sen University, Guangzhou 510275, China}
\def\SYUSFI{Sino-French Institute of Nuclear Engineering and Technology, Sun Yat-Sen University, Zhuhai, 519082, China}
\def\NKU{School of Physics, Nankai University, Tianjin 300071, China}
\def\YTU{Department of Physics,Yantai University, Yantai 264005, China}
\def\FDU{Key Laboratory of Nuclear Physics and Ion-beam Application (MOE), Institute of Modern Physics, Fudan University, Shanghai 200433, China}
\def\USST{School of Medical Instrument and Food Engineering, University of Shanghai for Science and Technology, Shanghai 200093, China}
\def\SJTUSC{Shanghai Jiao Tong University Sichuan Research Institute, Chengdu 610213, China}
\def\SPEIT{SJTU Paris Elite Institute of Technology, Shanghai Jiao Tong University, Shanghai, 200240, China}
\def\NNU{School of Physics and Technology, Nanjing Normal University, Nanjing 210023, China}
\def\SYUzhuhai{School of Physics and Astronomy, Sun Yat-Sen University, Zhuhai, 519082, China}

\author{Xuyang Ning}\affiliation{\shKeyLab}
\author{Abdusalam Abdukerim}\affiliation{\shKeyLab}
\author{Zihao Bo}\affiliation{\shKeyLab}
\author{Wei Chen}\affiliation{\shKeyLab}
\author{Xun Chen}\affiliation{\shKeyLab}\affiliation{\SJTUSC}
\author{Chen Cheng}\affiliation{\SYU}
\author{Zhaokan Cheng}\affiliation{\SYUSFI}
\author{Xiangyi Cui}\email[Corresponding author: ]{hongloumeng@sjtu.edu.cn}\affiliation{\TDLee}
\author{Yingjie Fan}\affiliation{\YTU}
\author{Deqing Fang}\affiliation{\FDU}
\author{Changbo Fu}\affiliation{\FDU}
\author{Mengting Fu}\affiliation{\pku}
\author{Lisheng Geng}\affiliation{\BUAA}\affiliation{\BUAALab}\affiliation{\zzu}
\author{Karl Giboni}\affiliation{\shKeyLab}
\author{Linhui Gu}\affiliation{\shKeyLab}
\author{Xuyuan Guo}\affiliation{\YaLongSD}
\author{Chencheng Han}\affiliation{\TDLee} 
\author{Ke Han}\affiliation{\shKeyLab}
\author{Changda He}\affiliation{\shKeyLab}
\author{Jinrong He}\affiliation{\YaLongSD}
\author{Di Huang}\affiliation{\shKeyLab}
\author{Yanlin Huang}\affiliation{\USST}
\author{Junting Huang}\affiliation{\shKeyLab}
\author{Zhou Huang}\affiliation{\shKeyLab}
\author{Ruquan Hou}\affiliation{\SJTUSC}
\author{Yu Hou}\affiliation{\MESJTU}
\author{Xiangdong Ji}\affiliation{\UMD}
\author{Yonglin Ju}\affiliation{\MESJTU}
\author{Chenxiang Li}\affiliation{\shKeyLab}
\author{Jiafu Li}\affiliation{\SYU}
\author{Mingchuan Li}\affiliation{\YaLongSD}
\author{Shuaijie Li}\affiliation{\TDLee}
\author{Tao Li}\affiliation{\SYUSFI}
\author{Qing Lin}\affiliation{\USTClab}\affiliation{\USTCdep}
\author{Jianglai Liu}\email[Spokesperson: ]{jianglai.liu@sjtu.edu.cn}\affiliation{\shKeyLab}\affiliation{\TDLee}\affiliation{\SJTUSC}
\author{Congcong Lu}\affiliation{\MESJTU}
\author{Xiaoying Lu}\affiliation{\SDUdep}\affiliation{\SDUlab}
\author{Lingyin Luo}\affiliation{\pku}
\author{Yunyang Luo}\affiliation{\USTCdep}
\author{Wenbo Ma}\affiliation{\shKeyLab}
\author{Yugang Ma}\affiliation{\FDU}
\author{Yajun Mao}\affiliation{\pku}
\author{Yue Meng}\affiliation{\shKeyLab}\affiliation{\SJTUSC}
\author{Ningchun Qi}\affiliation{\YaLongSD}
\author{Zhicheng Qian}\affiliation{\shKeyLab}
\author{Xiangxiang Ren}\affiliation{\SDUdep}\affiliation{\SDUlab}
\author{Nasir Shaheed}\affiliation{\SDUdep}\affiliation{\SDUlab}
\author{Xiaofeng Shang}\affiliation{\shKeyLab}
\author{Xiyuan Shao}\affiliation{\NKU}
\author{Guofang Shen}\affiliation{\BUAA}
\author{Lin Si}\affiliation{\shKeyLab}
\author{Wenliang Sun}\affiliation{\YaLongSD}
\author{Andi Tan}\affiliation{\UMD}
\author{Yi Tao}\affiliation{\shKeyLab}\affiliation{\SJTUSC}
\author{Anqing Wang}\affiliation{\SDUdep}\affiliation{\SDUlab}
\author{Meng Wang}\affiliation{\SDUdep}\affiliation{\SDUlab}
\author{Qiuhong Wang}\affiliation{\FDU}
\author{Shaobo Wang}\affiliation{\shKeyLab}\affiliation{\SPEIT}
\author{Siguang Wang}\affiliation{\pku}
\author{Wei Wang}\affiliation{\SYUSFI}\affiliation{\SYU}
\author{Xiuli Wang}\affiliation{\MESJTU}
\author{Zhou Wang}\affiliation{\shKeyLab}\affiliation{\SJTUSC}\affiliation{\TDLee}
\author{Yuehuan Wei}\affiliation{\SYUSFI}
\author{Mengmeng Wu}\affiliation{\SYU}
\author{Weihao Wu}\affiliation{\shKeyLab}
\author{Jingkai Xia}\affiliation{\shKeyLab}
\author{Mengjiao Xiao}\affiliation{\UMD}
\author{Xiang Xiao}\affiliation{\SYU}
\author{Pengwei Xie}\affiliation{\TDLee}
\author{Binbin Yan}\affiliation{\shKeyLab}
\author{Xiyu Yan}\affiliation{\SYUzhuhai}
\author{Jijun Yang}\affiliation{\shKeyLab}
\author{Yong Yang}\affiliation{\shKeyLab}
\author{Yukun Yao}\affiliation{\shKeyLab}
\author{Chunxu Yu}\affiliation{\NKU}
\author{Ying Yuan}\affiliation{\shKeyLab}
\author{Zhe Yuan}\affiliation{\FDU} %
\author{Xinning Zeng}\affiliation{\shKeyLab}
\author{Dan Zhang}\affiliation{\UMD}
\author{Minzhen Zhang}\affiliation{\shKeyLab}
\author{Peng Zhang}\affiliation{\YaLongSD}
\author{Shibo Zhang}\affiliation{\shKeyLab}
\author{Shu Zhang}\affiliation{\SYU}
\author{Tao Zhang}\affiliation{\shKeyLab}
\author{Wei Zhang}\affiliation{\TDLee}
\author{Yang Zhang}\affiliation{\SDUdep}\affiliation{\SDUlab}
\author{Yingxin Zhang}\affiliation{\SDUdep}\affiliation{\SDUlab} %
\author{Yuanyuan Zhang}\affiliation{\TDLee}
\author{Li Zhao}\affiliation{\shKeyLab}
\author{Qibin Zheng}\affiliation{\USST}
\author{Jifang Zhou}\affiliation{\YaLongSD}
\author{Ning Zhou}\email[Corresponding author: ]{nzhou@sjtu.edu.cn}\affiliation{\shKeyLab}\affiliation{\SJTUSC}
\author{Xiaopeng Zhou}\affiliation{\BUAA}
\author{Yong Zhou}\affiliation{\YaLongSD}
\author{Yubo Zhou}\affiliation{\shKeyLab}
\collaboration{PandaX Collaboration}
\author{Liangliang Su}\affiliation{\NNU}
\author{Lei Wu}\email[Corresponding author: ]{leiwu@njnu.edu.cn}\affiliation{\NNU}
\noaffiliation

\date{\today}
\begin{abstract}
 We report a search for light dark matter 
  produced through the cascading decay of $\eta$ mesons, which are created as a result of inelastic collisions between cosmic rays and Earth's atmosphere. We introduces a new and general framework, publicly accessible, designed to address boosted dark matter specifically, with which a full and dedicated simulation including both elastic and quasi-elastic processes of Earth attenuation effect on the dark matter particles arriving at the detector is performed.  In the PandaX-4T commissioning data of 0.63 tonne$\cdot$year exposure, no significant excess over background is observed. 
 The first constraints on the interaction between light dark matter  generated in the atmosphere and nucleus through a light scalar mediator are obtained.
 The lowest excluded cross-section is set at $5.9 \times 10^{-37}{\rm cm^2}$ for dark matter mass of $0.1$~MeV$/c^2$ and mediator mass of 300~MeV$/c^2$ . The lowest upper limit of $\eta$ to dark matter decay branching ratio is $1.6 \times 10^{-7}$.
\end{abstract}

\maketitle
Plenty of evidences from the astrophysics and cosmology observations indicate the existence of dark matter (DM), but  its nature still remains unknown. Direct detection experiments are  carried out globally to search for the signals of DM scattering off normal matters, based on a new interaction beyond the Standard Model of particle physics.
Traditional searches focus on the DM halo near the solar system, assuming a local DM density of approximately 0.3~GeV/$c^2/\rm cm^3$. Strong constraints have been placed on DM with mass above 10~GeV$/c^2$~\cite{XENON:2018voc, PandaX-4T:2021bab, LZ:2022ufs}. However, for light DM with mass at MeV$/c^2$ scale in the halo, the kinetic energy is not large enough to overcome the detector threshold, and thus the sensitivity to light DM degrades significantly.  Light DM candidates have been acquiring more and more interest, and various theoretical and experimental researches show great potential from direct detection to explore the light DM parameter space~\cite{boosted1,boosted2,Ge,ADM_ori,cui,PROSPECT,CDEX:2022fig,PandaX:2022osq,Cappiello:2019qsw,Yin:2018yjn}.

Recently, an interesting generic source of light DM flux was proposed~\cite{ADM_ori}, where the coupling between DM and nucleons may enable some mesons to partially decay to DM. The mesons generated from inelastic cosmic ray collisions with the atmosphere can produce an energetic flux of light DM. This process can be viewed as a continuous cosmic ray beam dump. 
The mass of the meson does not have to converted entirely to the DM mass, so that the arising DM particle can have a kinetic energy in the hundred MeV range.
Such a benchmark model is the hadrophilic scalar model~\cite{hardro, ADM_ori, Wu, ADM_new}, where a light Dirac fermion DM interacts with quarks through a light scalar mediator. 
Once generated, the light DM particles need to travel through Earth to reach the DM detectors placed in the underground laboratories. 
Due to the same coupling with nucleons, the DM flux gets attenuated through scattering with the nucleus in Earth. 
 Previously, a cut-off at $\mathcal O(100)$ MeV {\color{black} was} applied on the DM kinetic energy to ensure the dominance of coherent elastic process in the attenuation calculation~\cite{boosted1,PROSPECT,cui}, but it inevitably {\color{black}caused} a big loss of the sensitivity of underground detectors to these boosted DM particles.
In this letter, we perform a sensitive search for the light atmospheric DM using the commissioning data of PandaX-4T experiment, where an improved simulation of Earth attenuation effect is performed with the quasi-elastic process of a light scalar mediator included for the first time.

The PandaX-4T experiment is located in the China Jinping Underground Laboratory (CJPL), 
which has an overburden of 6700 meters water equivalent and a cosmic ray muon flux of $2.0 \times 10^{-10} {\rm /cm^2/s}$~\cite{CJPL_intro,CJPL2_intro}.
A dual-phased cylindrical time projection chamber (TPC) is operated with 3.7 tonne xenon in the sensitive volume. Two arrays of 3-inch photomultiplier tubes (PMTs) are placed on the top and bottom of the TPC to collect the signals. A scattering event with xenon is recorded as a prompt scintillation signal ($S1$) and a delayed electroluminescence light signal from ionization electrons ($S2$), based on which the scattering position and deposited energy are further reconstructed. Signal response models are constructed based on NEST v2.2.1~\cite{NESTv2.2.1, Szydagis2021} with parameters fitted to low energy calibration data. A more detailed description of the PandaX-4T experiment is given in Ref.~\cite{PandaX-4T:2021bab,PandaX:2018wtu,PandaX-4T:2022ldq,PandaX:2022aac}. 

The hadrophilic scalar mediator model introduces a singlet scalar mediator $S$ and a Dirac fermion DM $\chi$. 
To satisfy existing constraints on the flavor-changing neutral currents, the scalar mediator only couples to the DM and a specific quark flavor (up-quark in this model)~\cite{Batell:2017kty}. Therefore, there are only four free parameters, the DM mass $m_{\chi}$, the mediator mass $m_{S}$, the couplings $g_{\chi}$ and $g_{u}$. The corresponding Lagrangian  reads as follow~\cite{ADM_ori,hardro}:
\begin{equation}
\mathcal{L} \supset-g_{\chi} S \bar{\chi}_{L} \chi_{R}-g_{u} S \bar{u}_{L} u_{R}+\text { h.c. . }
\end{equation}
Under this model, the atmospheric DM flux is generated mainly in a cascade decay of $\eta$ mesons via the scalar mediator $S$, $\eta\rightarrow \pi^0 S \rightarrow \pi^0 \chi \bar{\chi}$, and the $\eta$ mesons are produced by inelastic collision of cosmic rays with the atmosphere. Contributions from heavier mesons like $\eta^\prime$ or $K^{+}$ are relatively much smaller~\cite{ADM_new}. 
The energetic $\eta $ flux from cosmic ray collision is calculated through a Monte Carlo simulation with the CRMC package as implemented in Ref.~\cite{ADM_ori}. 

Here we consider the situation that mediator is produced on-shell with $2 m_{\chi} < m_{S}<m_{\eta}-m_{\pi}$, where $m_\eta$ and $m_\pi$ are the mass of $\eta$ and $\pi^0$ respectively. 
The branching ratio of $\eta$ meson decaying to mediator $S$ is a function of $g_u^2$ and $m_S$~\cite{hardro,ADM_ori}.
Currently there is no dedicated measurement for $\eta\rightarrow \pi^0+{\rm invisible}$ decay.
The branching ratio ${\rm BR}(\eta\rightarrow \pi^0 S)$ is constrained by the uncertainties of measurements of the known $\eta$ decays~\cite{pdg}.
Compared with strong bounds on $g_u$, the coupling $g_\chi$ is much less constrained so that the decay $S \to \chi\bar{\chi} $ can dominate the decays of S~\cite{Batell:2018fqo}. For simplicity and maximizing the sensitivity, we assume $BR(S \to \chi\bar{\chi})=1 $ in our study. 
With these considerations, the benchmark set of parameters is chosen as $m_{S}= 300~{\rm MeV}/c^2$ and ${\rm BR}(\eta\rightarrow \pi^0 S)=1\times10^{-5}$.

The atmospheric DM from energetic $\eta$ decay is strongly boosted  as compared to the halo DM, with a kinetic energy $T_{\chi}$ up to $\mathcal{O}({\rm GeV})$. 
The possible interaction between the fast-moving DM and nucleus along the traveling trajectory includes coherent elastic, quasi-elastic (QE) and deep inelastic scattering (DIS) processes similar to the neutrinos~\cite{Paschos:2001np,Casper:2002sd,GENIE:2021npt,Alvey:2022pad}.
Especially, in the QE process, a fast-moving DM would collide directly with the constituent nucleons, so that one or more nucleons get excited or are dislodged from the nucleus.
Theoretical calculations indicate that the scalar-mediated DM-nucleus interaction is dominated by the QE process when the momentum transfer $q$, or equivalently the DM kinetic energy $T_{\chi}$, is above roughly 0.2~GeV~\cite{Su:2022wpj}. 
In this work, we consider $T_{\chi}$ up to 1~GeV and include both the elastic and QE processes in the calculation of Earth attenuation effect. In the detector, since the signal we are searching for is related to the final state of xenon nucleus after scattering, only the elastic process is considered for the purpose of validity and conservation.

With a scalar mediator, the DM-nucleon scattering cross section is dependent on the momentum transfer, so it is useful to define a momentum-independent reference cross-section as follows~\cite{Wu}
\begin{equation}
\bar{\sigma}_{\mathbf{n}} \equiv \frac{[Z y_{S p p}+(A-Z) y_{S n n}]^{2} g_{\chi}^{2} \mu_{\mathbf{n}}^{2}}{A^{2} \pi\left(q_{0}^{2}+m_{S}^{2}\right)^{2}},
\end{equation}
where $Z$ is the number of protons, $(A-Z)$ is the number of neutrons, the reference momentum transfer is $q_{0}^{2}=\alpha^{2} m_{e}^{2}$, the effective scalar-nucleon couplings are $y_{S p p}=0.014 ~g_{u} m_{p} / m_{u}$ and $ y_{S n n}=0.012  ~g_{u} m_{n} / m_{u}$  with $m_p$, $m_n$, and $m_u$ for the masses of the proton, neutron, and up-quark, respectively. $\mu_{\mathbf{n}}$ is the reduced mass of DM and nucleon.
For the DM traveling through Earth or scatter with target xenon in the underground detector,
the differential cross-section of the DM-nucleus elastic scattering involving a light scalar mediator as a function of nuclear recoil energy $E_R$ is expressed as
\begin{equation}
\frac{\mathrm{d} \sigma_{\chi N}}{\mathrm{~d} E_{R}}=\frac{\bar{\sigma}_{\mathbf{n}}A^2}{E_{R}^{\max }}\left(\frac{\mu_N}{\mu_{\mathbf{n}}}\right)^2\left|F_{\mathrm{DM}}(q)\right|^{2}|F_{N}(q)|^{2},
\label{dsigma}
\end{equation}
where $\mu_N$ is the reduce mass of DM particle and the target nucleus,  $E_R^{\max}$ is the maximum nuclear recoil energy for a given DM kinetic energy,  $q=\sqrt{2m_N E_R}$ is the momentum transfer, and $m_N$ is the mass of target nucleus, $F_{N}$ is the nuclear form factor~\cite{cui,Lewin:1995rx}, and $F_{\rm DM}$ is the DM momentum-dependent form factor~\cite{Wu} that can be expressed as
\begin{equation}
\left|F_{\mathrm{DM}}(q)\right|^{2}=\frac{\left(4 m_{N}^{2}+q^{2}\right)\left(4 m_{\chi}^{2}+q^{2}\right)\left(m_{S}^{2}+q_{0}^{2}\right)^{2}}{16 m_{N}^{2} m_{\chi}^{2}\left(m_{S}^{2}+q^{2}\right)^{2}}.
\label{dmff}
\end{equation}

The flux of atmospheric DM is calculated by integrating over the total atmospheric height, which is uniformly distributed on Earth surface. The attenuation effect for the DM passing through Earth before reaching the detector can be simulated using the PandaX-specific Monte Carlo package~\cite{code} developed in Ref.~\cite{cui}, which implements the Jinping Mountain profile and simulates both the velocity loss and angular deflection  of elastic scattering along the DM trajectory.
Compared with Ref.~\cite{cui} where only the DM flux from above the detector is considered, we improve the simulation by including the flux below the detector coming from the bottom part of Earth. For small interaction cross-section, the arrival flux from the bottom
 is nearly equal to that from the top, but relatively more scattering steps shift the DM kinetic energy to lower region. 

 In addition, the QE process is introduced in this simulation. 
  For a QE process, a DM particle with incoming momentum $k$ scatters directly with a constituent nucleon. The process is expressed as:
\begin{equation}
    \chi(k)+A\left(p_A\right) \rightarrow \chi\left(k^{\prime}\right)+X(\rightarrow n+Y) ,
\end{equation}
where $k^{\prime}$ indicates the momentum of the outgoing DM particle, $n$ for the scattering nucleon and $Y=A-1$ for the residual nucleus.
The differential cross section is then expressed in terms of the kinetic energy $T_\chi^{\prime}$ and direction $\Omega$ of the outgoing DM particle, 
\begin{equation}
\frac{\mathrm{d} \sigma_{\mathrm{QE}}}{\mathrm{d} T_\chi^{\prime} \mathrm{d} \Omega}=Z \frac{\mathrm{d} \sigma_p}{\mathrm{~d} T_\chi^{\prime} \mathrm{d} \Omega}+(A-Z) \frac{\mathrm{d} \sigma_n}{\mathrm{~d} T_\chi^{\prime} \mathrm{d} \Omega},
\end{equation}
where the details of differential cross-section of nucleon $\mathrm{d} \sigma_p/(\mathrm{d} T_\chi^{\prime} \mathrm{d} \Omega)$ and $\mathrm{d} \sigma_n/(\mathrm{d} T_\chi^{\prime} \mathrm{d} \Omega)$ for proton and neutron are given in Ref.~\cite{Su:2022wpj}.
 Compared with coherent elastic scattering, in the QE process there is no nuclear form factor suppression for the high energy DM, but the nucleon number $A^2$ enhancement reduces down to $A$.
 For a certain incident energy at each scattering step, we sample the interaction types 
according to the 
cross-sections of elastic and QE processes. If QE scattering happens, the distributions of outgoing DM particle are sampled from the differential cross-section with respect to the kinetic energy and deflection angle.

\begin{figure}[htbp]
  \includegraphics[width=\columnwidth]{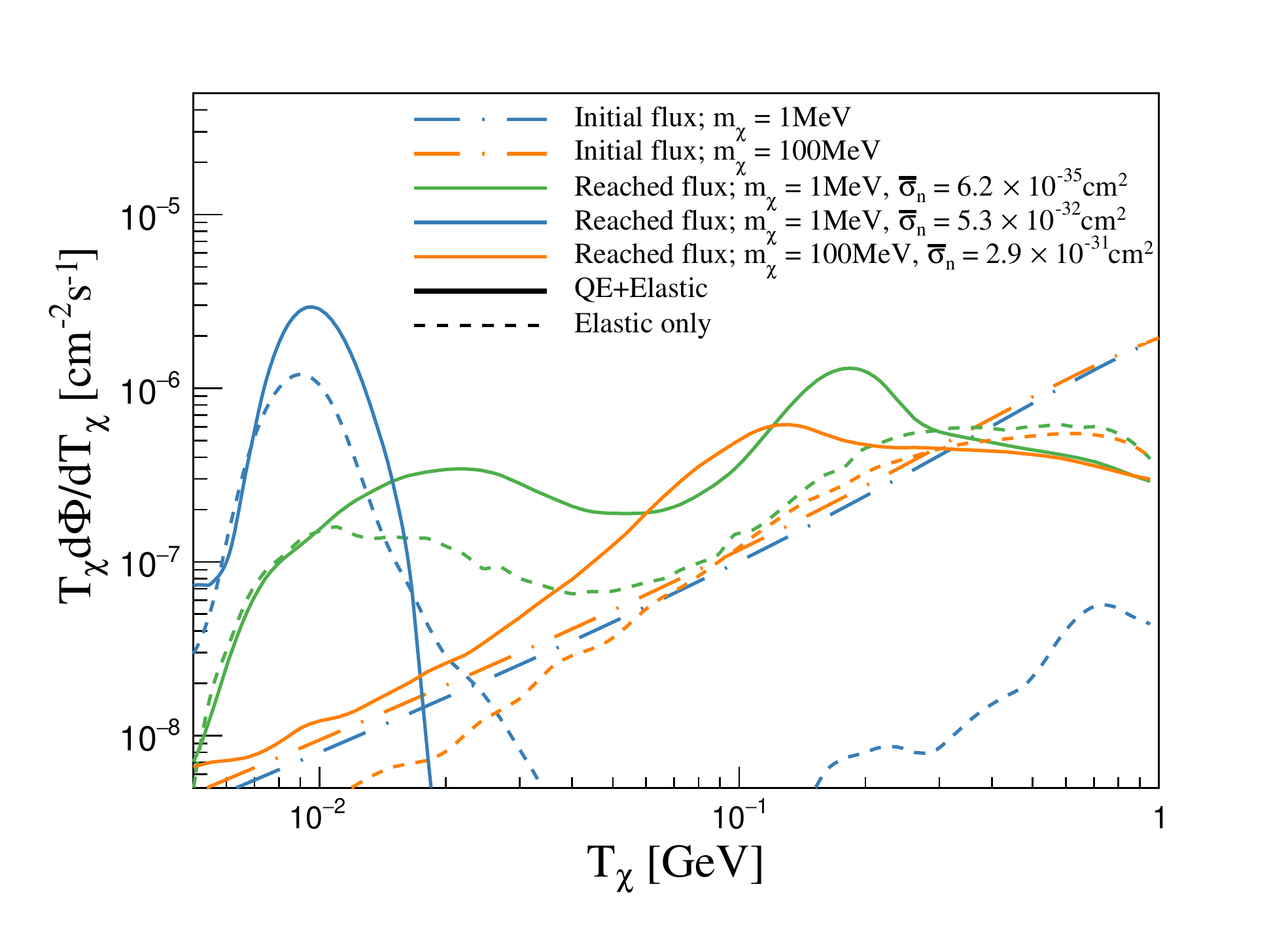}
  \includegraphics[width=\columnwidth]{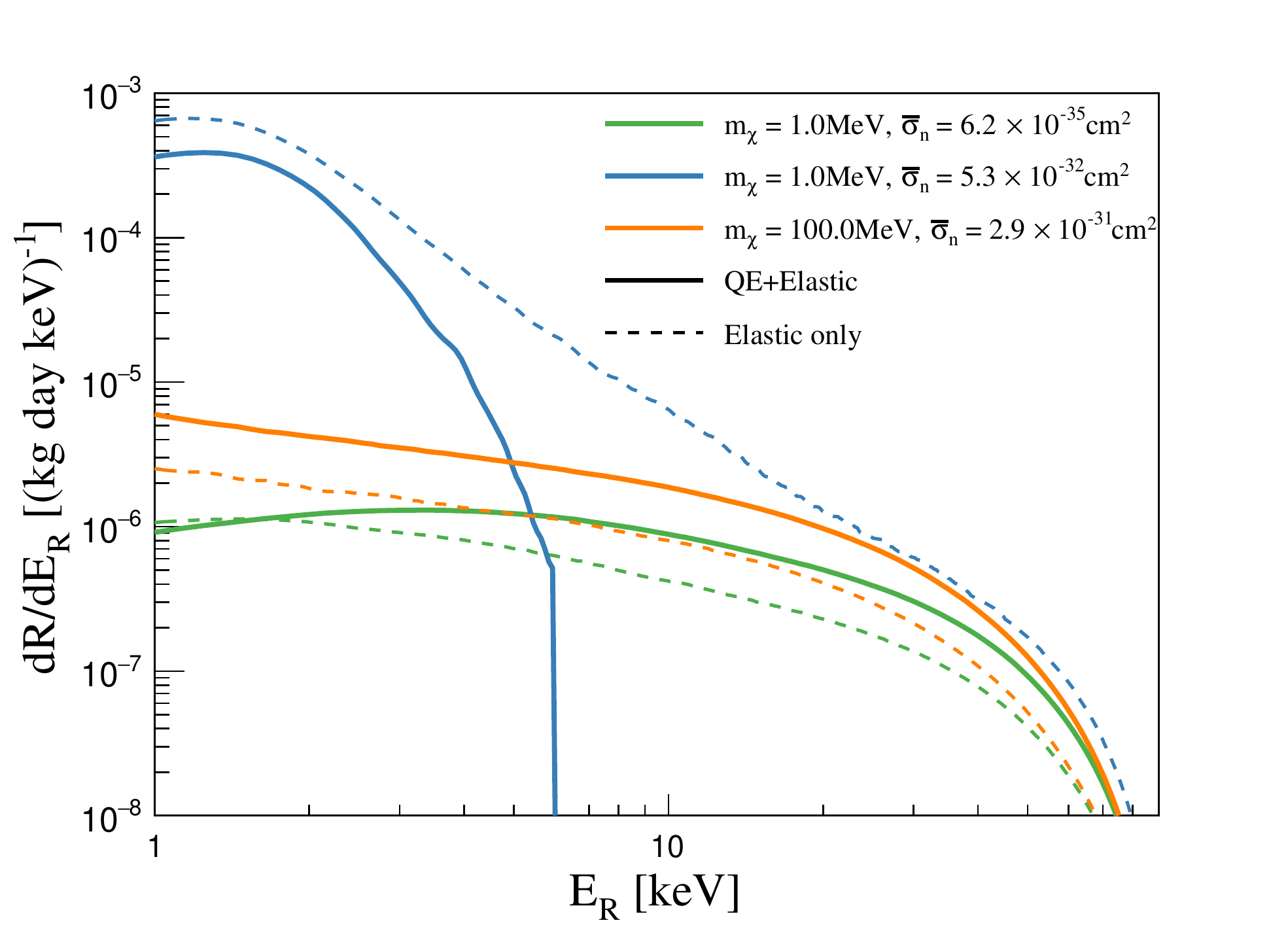}
  \caption{The upper panel shows the flux of atmospheric DM on Earth (dash-dotted lines) and that reaching the PandaX-4T detector (solid lines). The lower panel shows the differential event rate in xenon detector.
  For illustration, we take $m_{S}= 300~{\rm MeV}/c^2$ and ${\rm BR}(\eta\rightarrow \pi^0 S)=1\times10^{-5}$. The {\color{black}green and blue} lines are for DM mass of $m_\chi=1~{\rm MeV}/c^2$ with the reference cross-section $\bar{\sigma}_{\mathbf{n}}=6.2\times10^{-35}~{\rm cm^2} $ and $5.3\times10^{-32}~{\rm cm^2}$ respectively. The orange lines are for $m_\chi=100~{\rm MeV}/c^2$ with  $\bar{\sigma}_{\mathbf{n}}=2.9\times10^{-31}~{\rm cm^2}$.  
  The solid line shows the Monte Carlo simulation with quasi-elastic process included, while dashed line is with elastic-only assumption, for comparison. Quasi-elastic process not only reduces the reached flux, but also shifts the reached flux to the lower region. 
  }
  \label{fig:flux_ADM}
\end{figure}

{The upper panel of } Fig.~\ref{fig:flux_ADM} shows the atmospheric DM flux on Earth surface and that reaching the PandaX-4T detector after attenuation. 
Traveling through Earth would shift the DM flux reaching the detector to the lower kinetic energy region due to the velocity loss, which becomes quite obvious for large cross-sections. 
The dips in the flux reaching our detector near 60 MeV for $m_{\chi}=1$ MeV$/c^2$ is mainly due to the DM form factor, which enhances elastic scattering cross section for momentum transfer $q$ from a few tens to several hundreds MeV and causes a large energy loss~\cite{Wu}.
With the attenuated flux, the event rate of scattering off the xenon in the detector is shown in {the lower panel of Fig.~\ref{fig:flux_ADM}}. In calculating the xenon nuclear recoil signals, we consider DM particles with $T_\chi$ less than 1~GeV and elastic scattering process only as a conservative approach.  
For a comparison, the flux based on elastic-scattering-only assumption is overlaid, which indicates the importance of adding QE process in the attenuation calculation.

The data from the PandaX-4T commissioning run is used to search for this atmospheric DM, corresponding to 86.0 live-day exposure.
The data selection criteria follows Ref.~\cite{PandaX-4T:2021bab}, and the region of interest is defined with $S1$ from 2 to 135 $\rm PE$s and raw $S2$ from 80 to 20,000 $\rm PE$s. The background components include mainly tritium, ${}^{85}$Kr, $^{222}$Rn, material radioactivity, surface events, ${}^{136}$Xe, neutrons, neutrinos and accidental $S1-S2$ coincidence events, with detailed estimation described in Ref.~\cite{PandaX-4T:2021bab}. In total, 1058 events are selected in the data. A two-sided profile likelihood ratio method~\cite{Baxter:2021pqo} is adopted to test the signal hypothesis. We construct a standard unbinned-likelihood function~\cite{Cui2017, Wang2020} as
\begin{equation}
    \mathcal{L}_{\text{pandax}}=\left[\prod_{n=1}^{n_\mathrm{set}} \mathcal{L}_{n}\right] \times\left[\prod_{b} G(\delta_{b}, \sigma_{b})\right] \times \left[\prod_{p_{*}} G(\delta_{p_{*}}, \sigma_{p_{*}})\right],
\end{equation}
where $n_\mathrm{set} = 5$, the single set likelihood function $\mathcal{L}_{n}$ is defined as below
\begin{equation}
\begin{aligned}
    \mathcal{L}_{n} =& \operatorname{Poiss}(\mathcal{N}_{\text{obs}}^{n} \mid \mathcal{N}_{\text{fit}}^{n})\\ & \times \left[ \prod_{i=1}^{\mathcal{N}_{\text{obs}}^{n}}\frac{1}{\mathcal{N}_{\text{fit}}^{n}} \Big(N_{s}^{n} P_{s}^{n}(S1^{i}, S2^{i}_{\rm b}\vert\{p_{*}\})\right.\\
    & \left. + \sum_{b} N_{b}^{n}\left(1+\delta_{b}\right) P_{b}^{n}(S1^{i}, S2^{i}_{\rm b}\vert\{p_{*}\}) \Big)\right],
\end{aligned}
\end{equation}
where $\mathcal{N}_{\text{obs}}^{n}$ and $\mathcal{N}_{\text{fit}}^{n}$ are the total observed and fitted numbers of events for each data set $n$, respectively,
$N_s^n$ and $N_b^n$ are the number of DM signal and background events, $P_s^n(S1, S2_{\rm b})$ and $P_b^n(S1, S2_{\rm b})$ denote the two-dimensional PDFs. The systematic uncertainties of background estimation ($\sigma_b$) and nuisance parameters ($\sigma_{p_{*}}$) are constrained via Gaussian penalty terms $G(\delta, \sigma)$.

\begin{figure}[htbp]
  \centering
  \includegraphics[width=\columnwidth]{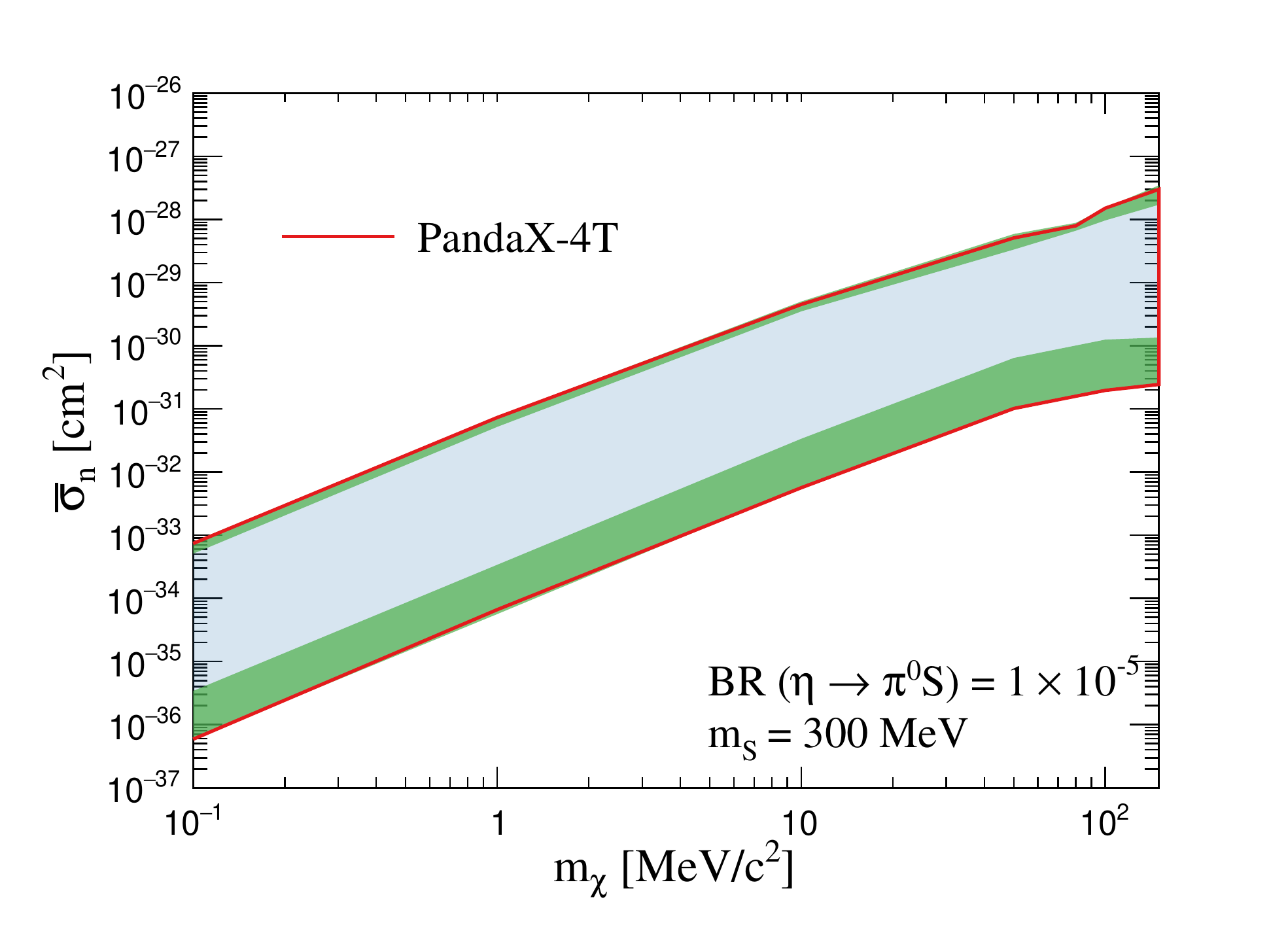}
  \includegraphics[width=\columnwidth]{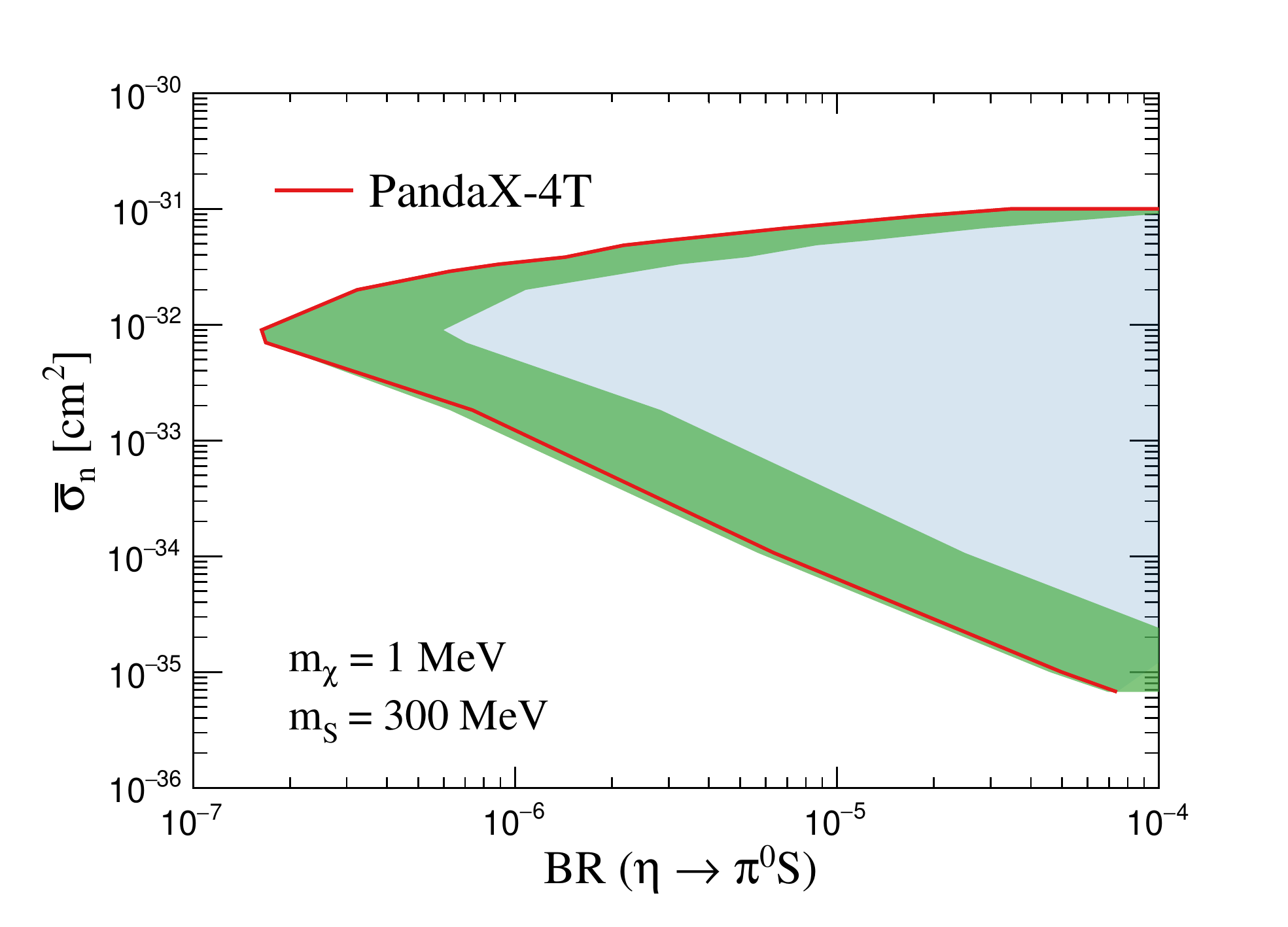}
  \caption{Top: 90\% C.L. excluded limit on $\sigma_{n}$ versus DM mass, with $m_{S}=300~{\rm MeV}/c^2$, ${\rm BR}\left(\eta \rightarrow \pi^{0} S\right)=1\times10^{-5}$. 
  Bottom: 90\% C.L. excluded limit on ${\rm BR}(\eta \rightarrow \pi^{0} S)$ versus $\sigma_{n}$, with $m_{S}=300~{\rm MeV}/c^2$, DM mass $m_{\chi}=1~{\rm MeV}/c^2$ and ${\rm BR}(S \rightarrow \chi\bar{\chi})=1$.
  The $\pm1\sigma$ sensitivity band is shown in green area. The region filled with blue color is excluded. 
  }
  \label{fig:limit_ADM}
\end{figure}

There is no significant excess observed in the data above the background under the hypothesis test. We derive 90\% confidence level (CL) constraints on the reference cross-section $\bar{\sigma}_{\mathbf{n}}$ versus DM mass $m_{\chi}$ for $m_{S}=300~{\rm MeV}/c^2$ and ${\rm BR}(\eta\rightarrow \pi^0 S)=1.0\times10^{-5}$, as shown in the upper panel of Fig.~\ref{fig:limit_ADM}.
The cut-off at $m_{\chi}=150~{\rm MeV}/c^2$ is due to the on-shell requirement of $m_{S}>2m_{\chi}$. 
The lower edge of the excluded band reaches $5.9 \times 10^{-37}~{\rm cm^{2}}$ at $m_{\chi}=0.1~{\rm MeV}/c^2$, and  $2.4 \times 10^{-31}~{\rm cm^{2}}$ at $m_{\chi}=150~{\rm MeV}/c^2$. 
The upper edge is $7.4 \times 10^{-34}~{\rm cm^{2}}$ at $m_{\chi}=0.1~{\rm MeV}/c^2$, and $3.0 \times 10^{-28}~{\rm cm^{2}}$ at $m_{\chi}=150~{\rm MeV}/c^2$, which indicates that the atmospheric DM  particles with  a too large scattering cross-section encounter very strong Earth attenuation and can hardly reach our detector. For light DM, the corresponding DM form factor in this model leads to an enhancement on the event rate of DM-nucleus scattering, which pushes the excluded region downward as compared to the conventional contact interaction.
 For smaller mediator mass $m_{S}$, the elastic scattering contribution becomes relatively larger~\cite{Su:2022wpj}, which results in a less {\color{black} loss on the kinetic energy} in Earth and pushes the upper edge of the exclusion band higher.

Alternatively, for a fixed DM mass, $m_\chi=1~{\rm MeV}/c^2$ for instance, the constraints can be converted into the $\eta$ meson decay branching ratio ${\rm BR}(\eta \rightarrow \pi^{0}S)$, as shown in the lower panel of Fig.~\ref{fig:limit_ADM}. The smallest upper limit on the branching ratio reaches $1.6 \times 10^{-7}$ at a reference cross-section of $9.0 \times 10^{-33}~{\rm cm^2}$.

\begin{figure}[htbp]
  \centering
  \includegraphics[width=\columnwidth]{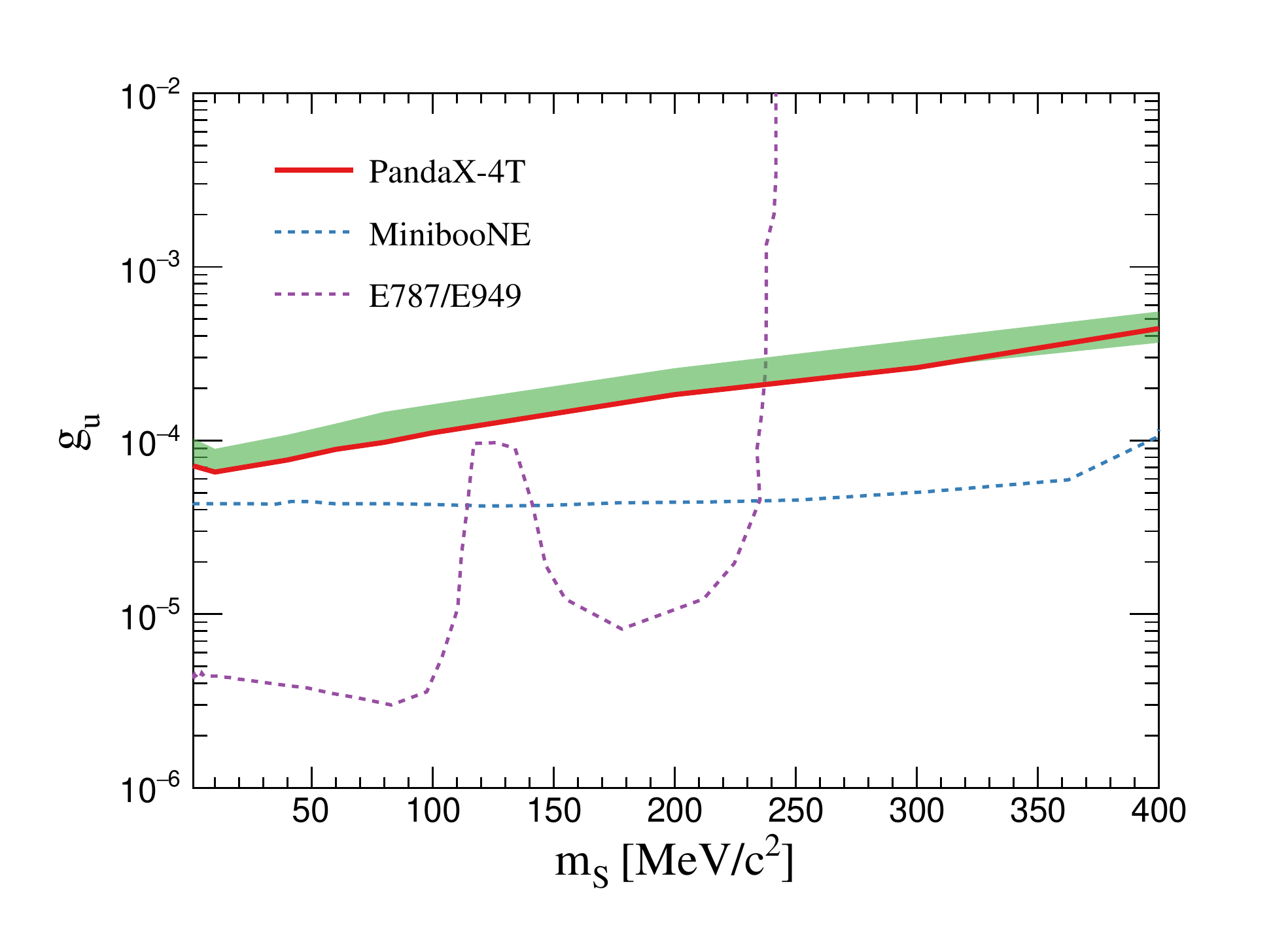}
  \caption{90\% C.L. excluded limit on $g_{u}$ versus mediator mass $m_{S}$, with DM mass $m_{\chi}=1/3\ m_{S}$ and $g_{\chi}=1$. $\pm1\sigma$ sensitivity band is shown in green area. Constraints derived by recasting results from MinibooNE and E787/E949 are taken directly from~\cite{hardro}.}
  \label{fig:limit_ADM_gums}
\end{figure}

To show the sensitivity of direct detection in testing the paradigm of a light dark sector with a mediator and sub-GeV DM, we give constrains on the mediator mass $m_{S}$ versus coupling $g_u$ in Fig.~\ref{fig:limit_ADM_gums}, by setting $g_{\chi}=1$ and DM mass $m_{\chi}= m_{S}/3$ as recommended in Ref.~\cite{hardro}.  Constraints on the coupling strength by recasting results from beam dump experiment MinibooNE~\cite{MiniBoo} and precision kaon measurement experiments E787 and E949~\cite{E787:1,E787:2,E949/787,E949:1} at Brookhaven are also shown for illustration~\cite{hardro}. Through searching for  the DM flux generated from cosmic ray, direct detection can provide comparable results on the light DM. 

In summary, we perform the first search for atmospheric DM 
using data from PandaX-4T commissioning run. For light DM, a dedicated calculation of Earth attenuation effect is done with both elastic and quasi-elastic scattering processes included. We demonstrate that quasi-elastic process is important in the evaluation of Earth attenuation effect, especially for those boosted DM particles.
With a scalar mediator $m_{S}=300$~MeV and BR$\left(\eta \rightarrow \pi^{0} S\right)=1.0\times10^{-5}$, we derive the strongest constraints on the reference DM-nucleon scattering cross-section. For DM mass $m_{\chi}=0.1$~MeV$/c^2$, the cross-section within  $5.9 \times 10^{-37} - 7.4 \times 10^{-34}~{\rm cm^2}$ is excluded. For DM $m_{\chi}=150$~MeV$/c^2$, the cross-section within $2.4 \times 10^{-31} - 3.0 \times 10^{-28}~{\rm cm^2}$ is excluded.
 We also derive upper limits on the BR$(\eta \rightarrow \pi^{0}S)$ with $m_{\chi}=1$~MeV$/c^2$, $m_{S}=300$~Me V$/c^2$ and BR$(S \rightarrow \chi\chi)=1$. The lowest upper limit of the branching ratio is $1.6 \times 10^{-7}$ for a reference cross section of $9.0 \times 10^{-33}~{\rm cm^2}$.
These results can be converted to the parameter space of $m_{S}$ versus $g_u$ of the hardrophilic DM model. For this model, the results from PandaX-4T direct detection experiment are 
comparable
to those from beam dump and precision meson measurement experiments.  PandaX-4T continues taking more physics data and is expected to improve the sensitivity by another order of magnitude with a 6-tonne-year exposure.


 
This project is supported in part by grants from National Science
Foundation of China (Nos. 12090061, 12005131, 11925502, 11835005), 
and by Office of Science and
Technology, Shanghai Municipal Government (grant No. 22JC1410100). We thank for the support from Double First Class Plan of
the Shanghai Jiao Tong University. We also thank the sponsorship from the
Chinese Academy of Sciences Center for Excellence in Particle
Physics (CCEPP), Hongwen Foundation in Hong Kong, Tencent
Foundation in China and Yangyang Development Fund. Finally, we thank the CJPL administration and
the Yalong River Hydropower Development Company Ltd. for
indispensable logistical support and other help.

\bibliographystyle{apsrev4-1}
\bibliography{refs.bib}


\end{document}


\begin{figure}[htbp]
  \includegraphics[width=0.8\columnwidth]{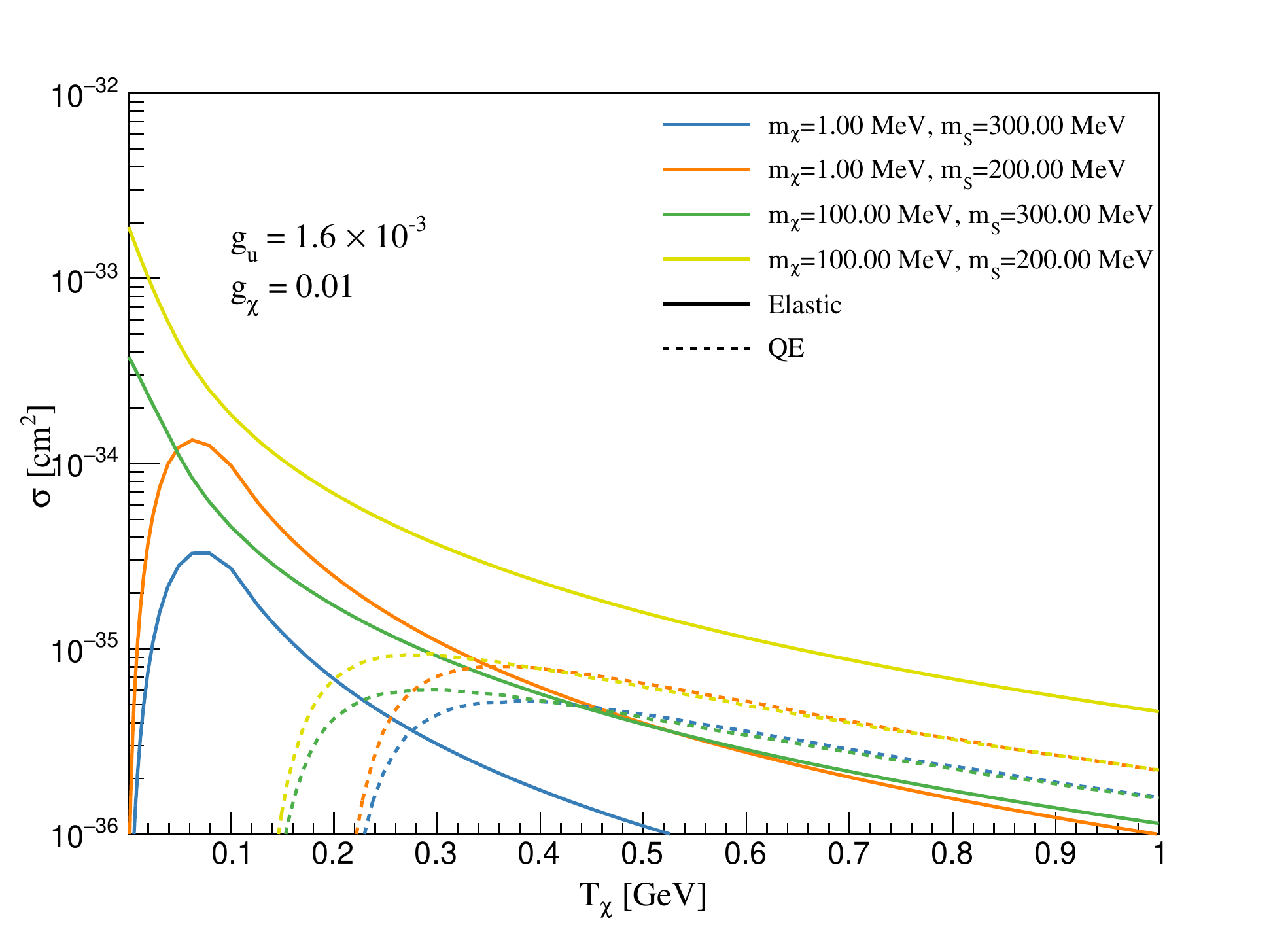}
  \caption{
  Comparison between elastic and Quasi-elastic cross sections used in this work. For illustration, Oxygen is taken as the target nucleus, $m_{\chi}$ is taken as 1 MeV and 100 MeV, $m_{S}$ is taken as 300 MeV for blue and green lines, and 200 MeV for orange and yellow lines. Solid lines are for elastic scattering process and dashed lines are for quasi-elastic scattering.
  }
  \label{fig:sigmaTot}
\end{figure}

\begin{figure}[htbp]
  \includegraphics[width=0.48\columnwidth]{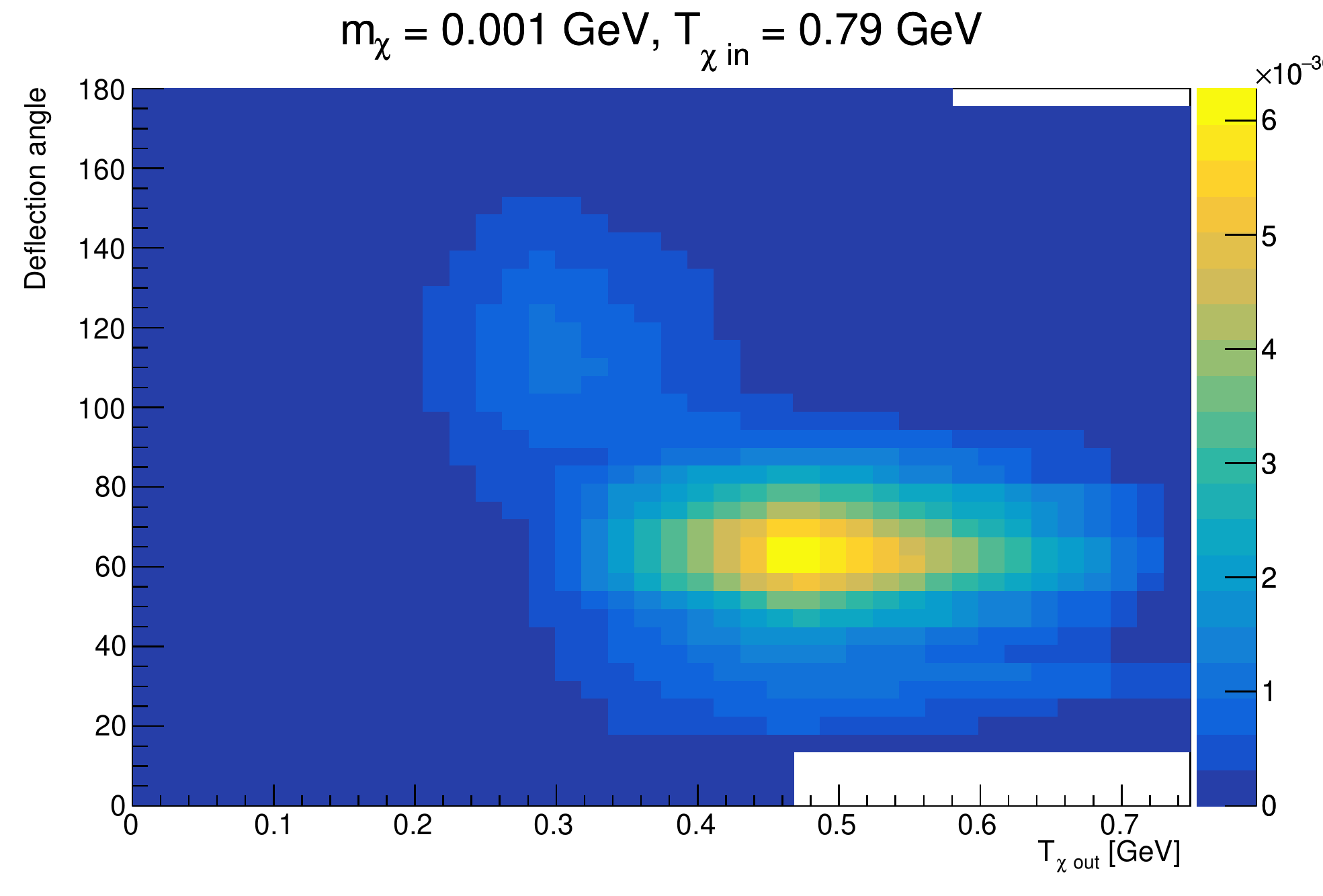}
  \includegraphics[width=0.48\columnwidth]{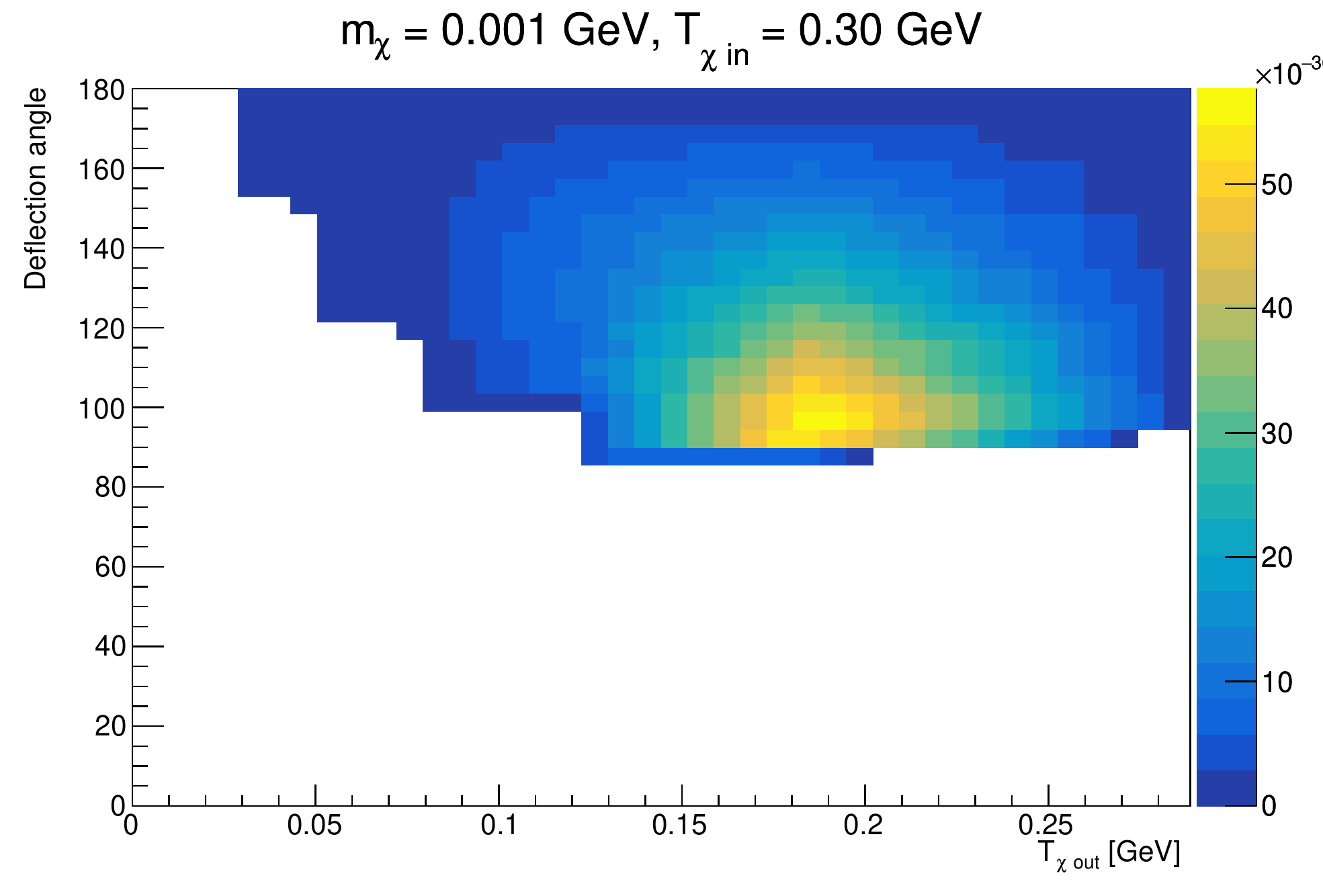}
  \caption{
  Two examples show the distribution of differential cross section with respect to $T_{\chi\ out}$ and deflection angle for QE process when DM scatter with Oxygen nucleus. $m_{\chi}$ is taken as 1 MeV and the DM incoming energy is taken as 0.79 GeV and 0.3 GeV, with $g_{u}=1.6 \times 10^{-3}$ and $g_{\chi}=0.01$ for illustration.
  }
  \label{fig:dsigma_QE}
\end{figure}